\documentclass[12pt]{iopart}
\usepackage{graphicx}
\usepackage{epsfig}
\usepackage{amssymb}
\providecommand{\openone}{\leavevmode\hbox{\small1\kern-3.8pt\normalsize1}}
\newcommand{\bra}[1]{\langle{#1} |}
\newcommand{\ket}[1]{| {#1} \rangle}
\newcommand{\mean}[1]{\langle {#1} \rangle}

\begin{document}

\title{Effects of low frequency noise cross-correlations in coupled superconducting 
       qubits}

\author{A. D'Arrigo, A. Mastellone, E. Paladino and G. Falci}
\address{MATIS CNR-INFM, Catania \&
Dipartimento di Metodologie Fisiche e Chimiche per l'Ingegneria,
Universit\`a di Catania, Viale Andrea Doria 6, 95125 Catania, Italy} 
\ead{darrigo@femto.dmfci.unict.it}

\begin{abstract}
We study the effects of correlated low frequency noise sources acting on
a two qubit gate in a fixed coupling scheme. A phenomenological
model for the spatial and cross-talk correlations is introduced. 
The decoherence inside the SWAP subspace is analyzed by combining analytic results based
on the adiabatic approximation and numerical simulations.
Results critically depend on amplitude of the low frequency noise
with respect to the qubits coupling strength. Correlations between noise
sources induce qualitative different behaviors depending on the
values of the above parameters.
The possibility to reduce dephasing due to correlated low frequency
noise by a recalibration protocol is discussed.
\end{abstract}

\pacs{03.67.Hk, 89.70.+c, 03.65.Yz, 03.67.Pp}

\maketitle

\section{Introduction}

Solid state nanodevices are at the forefront of present research
towards the implementation of quantum networks for quantum computation
and communication. The impressive development in device design and 
control tools achieved in recent years has by now to face intrinsic 
limitations due to material imperfections and fluctuations.
The resulting noise presents a variety of material and device
dependent features, ranging from noise spectra showing narrow resonances 
at selected frequencies (sometimes resonant with the nanodevice relevant 
energy scales) to low-frequency high-amplitude noise, often displaying 
1/f behavior. Modelling these fluctuations has naturally lead to overtake
the ubiquitous effective bath description via harmonic models and/or the 
hypothesis of linear coupling to the device under investigation. 

A typical example are background charge fluctuations which are known 
since more than 10 years to strongly affect the performance of Single-Electron 
Tunneling (SET) circuits~\cite{Zorin96}. Nowadays they represent the main limitation for
any nanocircuit gate requiring highly reliable electrostatic control.
This is clearly the case of charge~\cite{Nakamura99} and charge-phase~\cite{Vion02} 
superconducting qubits, but also of
semiconducting spin qubits elecrostatically coupled to form a two-qubit gate~\cite{Petta05}.     

The general belief is that background charge noise is due to the activity of
random traps for single electrons in dielectric materials surrounding the
island of SET devices or of superconducting nanocircuits. These traps may
have different trapping energies and switching times, $\gamma^{-1}$.
An ensemble of non interacting traps with a uniform distribution of trapping 
energies and a $1/\gamma$ distribution of switching rates may originate
the frequently observed 1/f noise~\cite{Weissman88}. Such a spectrum is indicative of numerous 
traps participating in the generation of the noise. On the other hand,
some samples clearly produce a telegraph noise with random switching between
a few states (with a magnitude of up to $0.1 \, e$ in SET devices)~\cite{Zorin96,Duty04}.
In addition, recent observations on superconducting qubits in different setups, 
have suggested the possibility that a few impurities may entangle with the 
device~\cite{Simmonds04,Ustinov07}. 
Such a variety of experimental facts may be consistently predicted by describing 
background charges as two-state systems whose dynamical behavior may turn from
quantum mechanical to classical with increasing temperature and/or with increasing
the strength of their dissipative interaction with the fluctuations of the surrounding
local host~\cite{Falci05}. Such a modellization clearly departs from the "conventional" bosonic bath
model and nicely predicts the different protocol-dependent decay laws  
of the coherent dynamics observed in charge and charge-phase qubits~\cite{Falci05}. Multiple
frequencies in the qubit dynamics and dependence on the uncontrollable impurities 
initial state at the beginning of measurement protocol are typical manifestation of 
non-gaussian character of background charge fluctuations~\cite{Paladino02}.

In order to limit the effect on single qubit gates of these material-specific 
fluctuations different strategies have been developed. Amongst the most successful
is the design of nanocircuits operating at "protected working points" 
insensitive to charge fluctuations to lowest order in the noise strength~\cite{Vion02}. 
Open- and closed-loop 
control protocols partly mutated from quantum optics and NMR~\cite{Slichter}, represent another
promising route~\cite{Ithier05,Nakamura02,Falci04}.

Presently, the effort of the scientific community working with Josephson qubits
is to appropriately extend the above strategies to  
multiqubit architectures, the first step being to
implement an efficient two-qubit gate. Different schemes to couple superconducting qubits
have been  proposed~\cite{coupling} and some experiments pointed out the
possibility to realize the desired entangled 
dynamics~\cite{Yamamoto02,Pashkin03,delft07}.
However, achievement of the needed high fidelity is still an ambitious task.
In addition to fluctuations experienced individually by each single qubit gate, 
coupled qubits, being usually built on-chip, may suffer from correlated noise due to
sources acting simultaneously on both sub-units.  
The effects on two-qubit gates  of uncorrelated and correlated bosonic baths
has been investigated~\cite{theory-twoqubit}. 
 
As far as the effect of background charges, opinions about the probable location of traps 
are divided and observations depend to a certain extent on the specific sample and on 
the junction geometry. However, there is unambiguous evidence that fluctuating traps 
located in the insulating substrate contribute essentially to the total noise in SET 
devices~\cite{Zorin96,zimerli}.
These traps are expected to induce similar fluctuations on the two islands built
on the same substrate. 
On the other hand fluctuating traps concentrated inside the oxide layer of the tunnel 
junctions, due to screening by the junction electrodes, are expected to act independently
on the two qubits~\cite{Zorin96}.

Fluctuating impurities acting simultaneously on coupled qubits represent a further
unconventional noise source which solid state nanodevices has to face~\cite{Hu07}. 
This is the subject of the present paper. 
Specifically, we will introduce a model for correlated charge noise on interacting
charge-phase qubits in a fixed capacitive coupling scheme. Relying on measurements
on SET circuits of power spectra on the two transistors and of the cross-spectrum power
density, we suppose a 1/f behavior for both the two channels spectra and the 
cross-spectrum~\cite{Zorin96}.  
In addition the cross-talk between the two qubits due to the capacitive coupling itself
between the islands will be discussed. 
Our analysis is based on analytical results obtained within the adiabatic approximation 
for the 1/f noise and on the numerical solution of the stochastic Schr\"odinger equation. 
Solving the dynamics from short-to-intermediate time scales allows complete understanding 
of the effects of correlations. Our work extends the analysis of  Ref.~\cite{Hu07} 
which, being limited to the long-times behavior, misses relevant features occurring
in the short time domain.
We find that usually correlations induce a faster decay of the coherent dynamics
compared to the action of independent fluctuations. Nevertheless,
under realistic values of low-frequency noise amplitude, increasing the
degree of correlation may instead lead to longer decoherence times.
Finally, the possibility to reduce the effects of low-frequency correlated noise via
open-loop recalibration protocols is discussed.

The paper is organized as follows: in Section 2 we introduce the setup consisting in two
Cooper pair boxes coupled by a capacitor. The cross-talk effect
and the charge noise sources responsible for correlations will be described  and their
correlation factor defined. In Section 3 we present a possible model for correlated
noise exhibiting 1/f power spectrum and cross-spectrum. In Section 4 relevant dynamical quantities 
for a two-qubit gate are introduced and analytic/numerical methods are illustrated. 
Section 5 includes our results for the entangled qubits dynamics in the presence of correlations. 
Conclusions are drawn in Section 6.    

\section{Coupled Cooper-pair boxes and noise correlations}
\label{section:Hamiltonian and noise correlation}
\begin{figure}[!th]
\centering
\includegraphics[width=100mm]{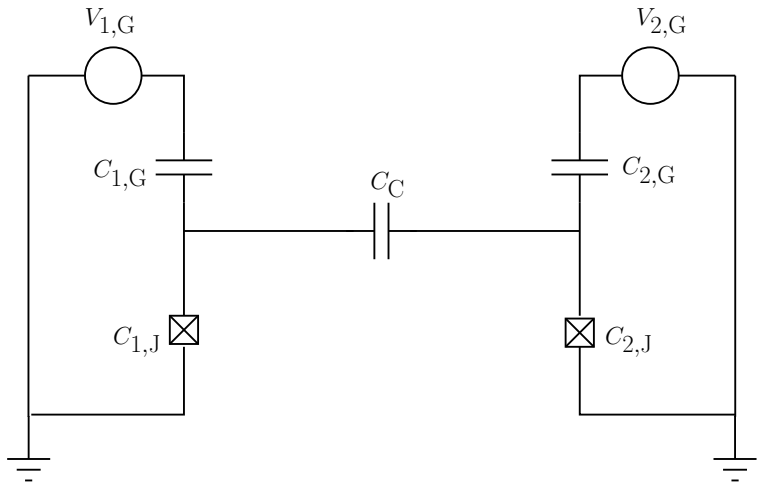}
\caption{Two CPB connected by a fixed capacitor.}
\label{fig:coupled-cpb}
\end{figure}

In the fixed capacitive coupling scheme for charge~\cite{Nakamura99} or charge-phase~\cite{Vion02} qubits 
the islands of two Cooper pair boxes (CPB) are connected through a 
capacitance~\cite{Yamamoto02,Pashkin03}, as
illustrated in \fref{fig:coupled-cpb}.
The system is described by the Hamiltonian
\begin{equation}
 {\cal H}_0= \sum_{\alpha \in \{1,2\}} {\cal H}_{\alpha}
   + E_{\mathrm{CC}}(\hat{q}_1-q_{1,\mathrm{x}})(\hat{q}_2-q_{2,\mathrm{x}}),
 \label{eq:Noiseless-TwoChargeQubit-H}
\end{equation}
where each CPB is modelled by
\begin{equation}
 {\cal H}_{\alpha}=[E_{\alpha,\mathrm{C}}
     (\hat{q}_\alpha-q_{\alpha,\mathrm{x}})^2 \,+\, 
     E_{\alpha,\mathrm{J}}\cos{\hat{\varphi}_\alpha}] \, .
 \label{eq:Noiseless-SingleQubit-H}
\end{equation}
$E_{\alpha,\mathrm{C}}=2e^2/C_{\alpha,\Sigma}$ 
is the charging energy of the island belonging to CPB $\alpha$,
the total island capacitance $C_{\alpha,\Sigma} = C_{\alpha,G} +
C_{\alpha,J}$ being the sum of the gate and junction capacitances.
$q_{\alpha,\mathrm{x}}=C_{\alpha,\mathrm{G}} V_{\alpha,\mathrm{G}}/(2e)$ 
is the corresponding dimensionless gate charge. 
Cooper pair tunneling across the Josephson junction $\alpha$
requires an energy  $E_{\alpha,\mathrm{J}}$.
$E_{\mathrm{CC}} =(2e)^2 C_\mathrm{T} / (C_{1,\Sigma} C_{2,\Sigma})$ 
is the coupling energy, with 
$1/C_\mathrm{T} = 1/C_\mathrm{C} + 1/C_{1,\Sigma} + 1/C_{2,\Sigma}$
the total inverse capacitance of the device.

The dimensionless charge $\hat{q}$ and the phase $\hat{\varphi}$ of each box
are conjugated variables,
$[ \hat{\varphi},\hat{q}] = \rmi$.
The system is subject to fluctuations of different origin. 
In part they arise from the control circuitry and can be described by an effective
impedance modelled by a conventional bosonic bath.
Noise sources of  microscopic origin are atomic defects located in the oxide of the tunnel
junctions, leading to fluctuations of the Josephson energy and  background charges
acting like additional uncontrollable $q_{\alpha,\mathrm{x}}$ sources. 
Devices based on the charge
variable are particularly sensitive to background charge fluctuations. 
Usually they can be modelled
as two-level fluctuators (TLF) inducing a bistable polarization of the superconducting island.  
A collection of TLF produces a noise whose spectral density  approximately follows a 1/f law.
They are responsible for the sensitive initial reduction of the amplitude
of coherent oscillations in single qubit gates observed when repeated measurements are 
performed~\cite{Falci05,Ithier05}. 
Fluctuations of polarization islands are expected 
to be a severe hindrance for coupled qubits gates based on the charge variable~\cite{Vion07}.
Low frequency charge fluctuations lead to an  additional stray contribution to 
the gate charge $q_{\alpha,\mathrm{x}}$, which can be modelled by a random variable  $\delta q_{\alpha,\mathrm{x}}(t)$
leading to
\begin{eqnarray}
 \fl {\cal H}= {\cal H}_0\, +\, \delta {\cal H}, 
  \label{eq:Noisy-TwoChargeQubit-H}\\
  \!\!\!\!\!\!\!\!\! \delta {\cal H}\,=\, 
    -\hat{q}_1 \big[2E_{1,\mathrm{C}}\delta q_{1,\mathrm{x}}(t)+
       E_{\mathrm{CC}}\delta q_{2,\mathrm{x}}(t)\big]  \,-\,
     \hat{q}_2 \big[E_{\mathrm{CC}}\delta q_{1,\mathrm{x}}(t)+
      2E_{2,\mathrm{C}}\delta q_{2,\mathrm{x}}(t)\big]
    \nonumber
\end{eqnarray}
Note that the coupling capacitance induces a cross-talk between the two devices, i.e.
fluctuations $\delta q_{1,\mathrm{x}}(t)$ acting on $\hat q_2$ and vice-versa.
As we already mentioned, background charges responsible for gate charge fluctuations
are spatially distributed in a device-dependent unpredictable way. 
Possibly they are partly located in the substrate,
partly in the oxide layer covering all electrodes, partly in the oxide barriers of
the tunnel junctions. Due to the shielding by the electrodes, impurities within 
tunnel junction $\alpha$ are expected to induce only gate charge fluctuations 
$\delta q_{\alpha,\mathrm{x}}(t)$. 
On the other hand, random arrangement of noise sources
in the bulk substrate originate correlations between gate charge
fluctuations to an extent depending on their precise location~\cite{Zorin96}.
\begin{figure}[!th]
\centering
\includegraphics[width=140mm]{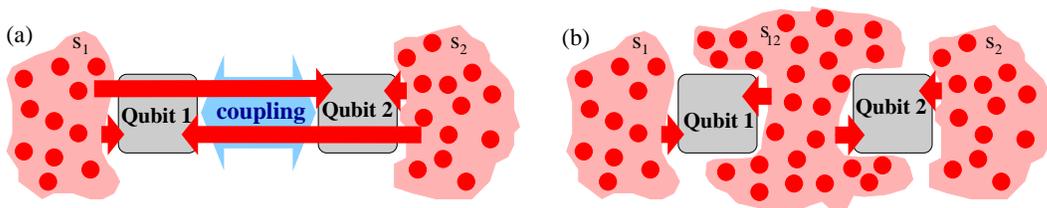}
\caption{Pictorial representation of  cross-talk (a) and spatial correlations (b). 
(a) The uncorrelated noise sources $s_1$ and $s_2$ act on coupled qubits: because of the coupling each 
qubit also suffers from the noise source directly acting on the other qubit. 
(b) Two non interacting qubits in the presence of  $s_1$, $s_2$ and of
$s_{12}$ that simultaneously acts on both qubits. 
$s_{12}$ may represent a set of  of impurities
located in the insulating substrate. It generates a  fluctuating interaction
even in the absence of direct coupling between the two qubits.}
\label{fig1}
\end{figure}
Pictorially, we may  separate impurities 
in two ensembles $\{s_1,s_2\}$ and $s_{12}$ influencing either each sub-unit
or both islands, as illustrated in \fref{fig1}. 
The noise $\delta q_{\alpha,\mathrm{x}}$ viewed by 
qubit $\alpha$ is due both to $s_\alpha$ and $s_{12}$. Correlations between
 $\delta q_{1,\mathrm{x}}$ and  $\delta q_{2,\mathrm{x}}$ originate from
set  $s_{12}$, are termed \textit{spatial correlations}.

In the above phenomenological description,
we assume that $\delta q_{\alpha,\mathrm{x}}(t)$ are stationary stochastic processes
having  zero average and the same variance $\overline{\sigma}^2$.  
We quantify the degree of spatial correlation between $\delta q_{1,\mathrm{x}}$ and 
$\delta q_{2,\mathrm{x}}$ via 
$\langle \delta q_{\alpha,\mathrm{x}}(t) \delta q_{\beta,\mathrm{x}}(t)\rangle = 
[\delta_{\alpha\beta}+\mu_{\mathrm{sp}}(1-\delta_{\alpha\beta})]\overline{\sigma}^2$.
Evaluation of $\mu_{\mathrm{sp}}$ would require a microscopic description of
the device and it is expected to depend on the
dimension of the boxes and on their relative distance, on the
specific spatial distribution of impurities $s_{12}$ 
and on the relative weights on $\delta q_{\alpha,\mathrm{x}}(t)$ of fluctuations due to set 
$s_{12}$ and $s_\alpha$, as shown in Ref.\cite{Zorin96}.

Cross-talk due to coupling and spatial correlations enter the overall noise $X_\alpha$ felt by each sub-unit:
\begin{eqnarray}
 \fl \qquad X_1(t) = 
     2E_{1,\mathrm{C}}\,\delta q_{1,\mathrm{x}}(t)\,
            +\,E_{\mathrm{CC}}\,\delta q_{2,\mathrm{x}}(t) 
\label{X1}\\    
 \fl\qquad X_2(t) = 
     E_{\mathrm{CC}}\,\delta q_{1,\mathrm{x}}(t)\,
            +\,2 E_{2,\mathrm{C}}\,\delta q_{2,\mathrm{x}}(t) 
\label{X2}
\end{eqnarray}
The amount of correlation between $X_1$ and $X_2$ 
may be quantified by a correlation coefficient, which is defined for general 
stochastic processes $\xi_1$ and $\xi_2$ as ~\cite{Papoulis}:
\begin{eqnarray}
   \mu=\frac{\langle [\xi_1(t)-\overline{\xi}_1][\xi_2(t)-\overline{\xi}_2]\rangle}
       {\sqrt{\langle[\xi_1(t)-\overline{\xi}_1]^2\rangle
              \langle[\xi_2(t)-\overline{\xi}_2]^2\rangle}} \, ,
\label{XCorrelatioFactor}
\end{eqnarray}
where $\langle \cdot \rangle $ indicates the ensemble average, 
and $\overline{\xi}_i \equiv \langle \xi_i(t)\rangle $.
>From \eref{X1} and \eref{X2} we obtain
\begin{eqnarray}
 \langle X_\alpha^2 \rangle &=&(4E_{\alpha,\mathrm{C}}^2+E_{\mathrm{CC}}^2
               +4\mu_\mathrm{sp} E_{\mathrm{\alpha,C}} E_{\mathrm{CC}} )\overline \sigma^2,
 \nonumber \\
 \langle X_1X_2 \rangle &=& [2E_{\mathrm{CC}}(E_{1,\mathrm{C}}+E_{\mathrm{2,C}})
  +\mu_\mathrm{sp}(4E_{1,\mathrm{C}}E_{2,\mathrm{C}}+E_{\mathrm{CC}}^2)]\overline \sigma^2 
 \nonumber 
\end{eqnarray}
thus the correlation coefficient of $X_1$ and $X_2$ reads
\begin{equation}
   \mu=\frac{2E_{\mathrm{CC}}(E_{1,\mathrm{C}}+E_{2,\mathrm{C}})+
       \mu_\mathrm{sp}(4E_{1,\mathrm{C}}E_{2,\mathrm{C}}+E_{\mathrm{CC}}^2)} 
        {\sqrt{(4E_{1,\mathrm{C}}^2+E_{\mathrm{CC}}^2+
            4\mu_\mathrm{sp}E_{1,\mathrm{C}}E_{\mathrm{CC}})
               (4E_{2,\mathrm{C}}^2+E_{\mathrm{CC}}^2+
            4\mu_\mathrm{sp}E_{2,\mathrm{C}}E_{\mathrm{CC}})}} \, .
\label{eq:TotalCorrelatioFactor}
\end{equation}
In the absence of spatial correlations $\delta q_{\alpha,\mathrm{x}}$ are independent
and only the effect of cross-talk is left. The correlation coefficient in this case reduces to
\begin{equation}
   \mu_{\mathrm{ct}}=
   \frac{2E_{\mathrm{CC}}(E_{1,\mathrm{C}}+E_{2,\mathrm{C}})} 
    {\sqrt{(4E_{1,\mathrm{C}}^2+E_{\mathrm{CC}}^2)
           (4E_{2,\mathrm{C}}^2+E_{\mathrm{CC}}^2)}} 
   \simeq
   \frac{4E_{\mathrm{C}}E_{\mathrm{CC}}} {(4E_{\mathrm{C}}^2+E_{\mathrm{CC}}^2)}.
\label{eq:CrossTalkingCorrelatioFactor}
\end{equation}
In the last approximation we have supposed  $E_{\alpha,\mathrm{C}} \simeq E_{\mathrm{C}}$,
within manufacture tolerances.
For typical values of parameters for charge qubits  $E_{\mathrm{C}} \gg E_{\mathrm{CC}}$
thus  $\mu_{\mathrm{ct}} \simeq E_{\mathrm{CC}}/E_{\mathrm{C}}$, 
giving values between 0.015 \cite{Vion07} and 0.12 \cite{Pashkin03}.
Clearly, larger values of the coupling strength $E_{\mathrm{CC}}$,
desirable to produce faster two-qubit gates, 
would also lead to higher cross-talk correlations $\mu_{\mathrm{ct}}$.
In general, the correlation coefficient \eref{eq:TotalCorrelatioFactor}
for $E_{\alpha,\mathrm{C}} \simeq E_{\mathrm{C}}$ is
approximately given by
\begin{equation}
   \mu \simeq 
     \frac{4E_{\mathrm{C}}E_{\mathrm{CC}}+
        \mu_\mathrm{sp}(4E_{\mathrm{C}}^2+E_{\mathrm{CC}}^2)}
        {4E_{\mathrm{C}}^2+E_{\mathrm{CC}}^2+
        4\mu_\mathrm{sp}E_{\mathrm{C}}E_{\mathrm{CC}}} \,=\,
     \frac{\mu_\mathrm{ct}\,+\,\mu_\mathrm{sp}} 
                   {1+\mu_\mathrm{ct}\mu_\mathrm{sp}}
   \label{eq:SympifiedTotalCorrelatioFactor}
\end{equation}
where we used \eref{eq:CrossTalkingCorrelatioFactor}.
Strong correlations between $X_1$ and $X_2$, $\mu \approx 1$, 
may originate either from large cross-talk or from large spatial correlations.
For instance by engeneering device design~\cite{You05} it could be possible
to get $\mu_\mathrm{ct} \simeq 1$ implying $\mu \simeq 1$.
On the other hand, in the presence of a low level of correlations
$\mu_\mathrm{sp}\mu_\mathrm{ct} \ll 1$, equation \eref{eq:SympifiedTotalCorrelatioFactor} 
simplifies to  $\mu \simeq \mu_\mathrm{sp} + \mu_\mathrm{ct}$.

In general \eref{eq:TotalCorrelatioFactor} gives the overall amount of correlation between
fluctuations affecting the two CPBs.
In the following we will not specify the physical mechanism responsible for these correlations.
We will simply suppose the existence of a degree of correlation between the fluctuations
$X_1$ and $X_2$ quantified by the coefficient $\mu$.

\section{Charge noise power spectra and cross-spectrum}
\label{section:Noise-model}
Measurements of charge noise due to background charge fluctuations in SET devices~\cite{Zorin96} have
revealed a 1/f behavior at low frequencies (measurements extend down to about 1 Hz), 
with a roll-off frequency of $100~-~1000$ Hz. 
Two SET whose islands are positioned about $100$ nm apart show a similar 1/f behavior for the cross-spectrum 
(defined later by \eref{SX12}) indicating correlations 
between fluctuations affecting both islands.
Similarly, measurements of charge noise in charge-phase~\cite{Ithier05} qubits shows a 1/f 
behavior for $f < 100$ kHz whose amplitude depends on temperature, on junction size and
on screening of the island by electrodes. Echo experiments suggest that 1/f noise
extends up to $1$ MHz.
In this setup charge noise at higher frequencies 
(up to $10$ MHz) is due to driving and readout subcircuits and it is characterized by white spectrum.
Measurements of energy relaxation processes in charge qubits have suggested that charged
impurities may also be responsible for ohmic noise at GHz frequencies~\cite{Astafiev04}.
To our knowledge, measurements of cross-spectrum 
on these class of nanodevices have not been reported in the literature. 
It is however expected that, similarly to SET devices, correlations between fluctuations
acting on superconducting islands of the two on-chip CPBs display 1/f cross-spectrum at low
frequencies.

Our goal is to introduce a model for the fluctuations $X_1(t)$ and $X_2(t)$
such that both power spectra and cross-spectrum 
\begin{eqnarray}
 S_{X_\alpha}(\omega)\,&=&\,\int_{-\infty}^{+\infty} \rmd \tau\, \rme^{-\rmi \omega \tau}\, 
\, [\,\langle X_\alpha(t+\tau)X_\alpha(t)\rangle -  \overline{X}_\alpha^2\,] 
\label{SXalpha} \\
S_{X_1 X_2}(\omega) \,&=&\,\int_{-\infty}^{+\infty} \rmd \tau\, 
    \rme^{-\rmi \omega \tau}\, [\, \langle X_1(t+\tau)X_2(t)\rangle - 
                    \overline{X}_1\overline{X}_2\ \,]
\label{SX12}
\end{eqnarray}
display similar 1/f behavior at low frequencies and are characterized by 
a finite correlation coefficient defined by \eref{XCorrelatioFactor}.
To this end we introduce two {\it independent} stationary stochastic processes,
$n_1(t)$ and $n_2(t)$  
with the same average and characterized by the same  autocovariance function
and spectrum~\cite{Papoulis}
\begin{eqnarray}
 C_{n_\alpha n_\alpha}(\tau) \,&=& \, 
		\langle n_\alpha(t+\tau)n_\alpha(t)\rangle -  \overline{n}_\alpha^2 
		\equiv \,C(\tau)\nonumber\\
 S_{n_\alpha n_\alpha}(\omega)\,&=&\,
		\int_{-\infty}^{+\infty} \rmd \tau\, \rme^{-\rmi \omega \tau}\, C(\tau)
 		\equiv \, S(\omega) \, .
\label{PSD}
\label{Autocovariance}
\end{eqnarray}
The processes $X_1(t)$ and $X_2(t)$, defined as linear combinations of $n_1(t)$ and  $n_2(t)$
\begin{eqnarray}
   X_1(t)&=& \sqrt{1-\eta}\,n_1(t)\, +\, \sqrt{\eta}\,n_2(t) \nonumber\\
   X_2(t)&=& \sqrt{\eta}\,n_1(t)\, +\, \sqrt{1-\eta}\,n_2(t) \, ,
\label{CorrelatedSP}
\end{eqnarray}
with $\eta \in [0,\frac{1}{2}]$, are correlated and
their correlation coefficient reads
\begin{equation}
   \mu=2\sqrt{\eta(1-\eta)} \,.
\label{CorrelationFactor}
\end{equation}
Thus $\mu$ is a monotonic function of $\eta \in [0,\frac{1}{2}]$ ranging in the interval $[0,1]$. 
If $\eta=0$, $X_1$ and $X_2$ reduce respectively to the uncorrelated processes $n_1$ and $n_2$
and $\mu=0$.
Instead when $\eta=\frac{1}{2}$ the correlation factor reaches its maximum value $\mu=1$.
In this case  $X_1$ and $X_2$ reduce to the same process: 
\begin{equation}
 \mu=1 \quad \Rightarrow \quad  X_1(t)\,=\,X_2(t)\,= 
                   \frac{1}{\sqrt{2}}[n_1(t)+n_2(t)] \, .
\end{equation}
The autocovariance functions of $X_1(t)$ and $X_2(t)$ are identical and read
\begin{equation}
   C_{X_\alpha X_\alpha}(\tau) = \langle X_\alpha(t+\tau) X_\alpha(t)\rangle -  \overline{X}_\alpha^2= \,
                     (1-\eta)\,C_{n_1n_1}\,+\, \eta\, C_{n_2n_2} = C(\tau) \, ,
\label{Autocovariance_xx}
\end{equation}
therefore $X_1$ and $X_2$ have the same variance $\sigma^2$ and equal power spectra 
$S_{X_1}(\omega)=S_{X_1}(\omega)=S(\omega)$ given by \eref{PSD}. 
This simple model for correlated noise allows, by changing the arbitrary parameter $\eta$, to 
modulate the correlation coefficient between $X_1$ and $X_2$ 
from 0 to 1, maintaining the desired spectrum $S(\omega)$ for both processes.
It is worth noticing that the first order statistics of $X_1$ and $X_2$ depends on
$\eta$: $\overline{X}_1=\overline{X}_2=\overline{n}(\sqrt{1-\eta}+\sqrt{\eta})$.
To avoid this dependence we set $\overline{n}=0$, implying vanishing  average values for 
$X_1$ and $X_2$.
The correlation factor enters the  cross-covariance and the cross-spectrum
of $X_1$ and $X_2$~\cite{Papoulis} 
\begin{eqnarray}
   C_{X_1 X_2}(\tau) \, &=& \, \langle X_1(t+\tau)X_2(t)\rangle - 
                    \overline{X}_1\overline{X}_2\,= \,
       \mu \,C(\tau) \, ,
\label{Cross-covariance}
\\
   S_{X_1 X_2}(\omega) \, &= & \,\int_{-\infty}^{+\infty} \rmd \tau\, 
    \rme^{-\rmi \omega \tau}\, C_{X_1 X_2} (\tau) \,=\, \mu\, S(\omega) \, .
\label{Cross-Spectrum}
\end{eqnarray}
It can therefore be detected by spectral analysis via~\cite{Bendat} 
\begin{equation}
   \frac{S_{X_1 X_2}(\omega)}{\sqrt{S_{X_1}(\omega)S_{X_1}(\omega)}}\,=\,\mu \, .
\label{Spectrum-correlation factor}
\end{equation}
By measuring power spectra and cross-spectrum of  voltage fluctuations across each SET  
in the frequency range $1$ to $10$ Hz, Zorin et al. estimated 
according to \eref{Spectrum-correlation factor} the correlation coefficient 
$\mu=0.15 \pm 0.05$~\cite{Zorin96}.  

In order to obtain a 1/f spectrum for processes $X_\alpha(t)$ we adopt
a commonly employed model which consists of an ensemble of independent TLF. 
Each fluctuator incoherently switches between two metastable 
levels, with a rate $\gamma_k$, producing a random signal $\xi_k(t)$. This
signal has a lorentzian power spectrum,
$S_{\xi_k}(\omega)=\frac{1}{2}v_k^2 \gamma_k/(\gamma_k^2+\omega^2)$,
$v_k$ being the difference between the values assumed by  $\xi_k(t)$.
When the switching rates $\gamma_k$ are distributed according to
$P(\gamma) \propto 1/\gamma$ in $[\gamma_\mathrm{m},\gamma_\mathrm{M}]$,
the overall noise obtained by summing all TLFs contributions displays
a $1/\omega$ behavior in  $[\gamma_\mathrm{m},\gamma_\mathrm{M}]$ \cite{Weissman88}
\begin{equation}
 \xi=\sum_k \xi_k(t) \quad \Rightarrow \quad 
  S_{\xi}(\omega)=\sum_{k=1}^{N_{\mathrm{TLF}}} \frac{v_k^2 \gamma_k}{2(\gamma_k^2+\omega^2)}
  \simeq \frac{\cal A}{\omega} 
\label{eq:1_f_model}
\end{equation}
where ${\cal A}=\pi\mean{v^2}N_{\mathrm{TLF}}/[4\ln(\gamma_\mathrm{M}/\gamma_\mathrm{m})]$ and $N_{\mathrm{TLF}}$ is
the total number of fluctuators. 
If the  independent random processes $n_1(t)$
and $n_2(t)$ are generated as a  sum of such an ensemble of TLFs,
the spectrum of each $n_\alpha(t)$ will be 1/f in  $[\gamma_\mathrm{m},\gamma_\mathrm{M}]$
and will have  variance 
$\sigma^2=\frac{1}{2\pi}\int d\omega S(\omega)= \frac{1}{4}N_{\mathrm{TLF}}\mean{v^2}$. 
The 1/f correlated stochastic processes $X_1(t)$ and $X_2(t)$ are obtained 
from \eref{CorrelatedSP} once the phenomenological correlation factor $\mu$
is fixed.

\section{Two qubit gate and relevant dynamical quantities}
\label{subsection:Two Qubit Hamiltonian}
At sufficiently low temperatures each CPB may operate as an effective two-state system, 
the coupled boxes implementing a two qubit gate. In the fixed coupling scheme
the interaction is switched on by individually manipulating each qubit to
enforce mutual resonance conditions~\cite{Yamamoto02,Pashkin03}. This allows
the realization of elementary two qubit operations.
We denote the lowest eigenstates  of 
each CPB as $\{\ket{+}_\alpha,\ket{-}_\alpha\} $,
with splitting depending on the control parameter $q_{\alpha,\mathrm{x}}$,
$\Omega_\alpha(q_{\alpha,\mathrm{x}})$.
By operating at the so called ``charge protected point'', $q_{\alpha,\mathrm{x}}=1/2$,
the system is insensitive to charge fluctuations at lowest order, meaning
that $d\Omega_\alpha(q_{\alpha,\mathrm{x}})/dq_{\alpha,\mathrm{x}}|_{q_{\alpha,\mathrm{x}}=0.5} =0$
\cite{Vion02}.
In a pseudo-spin description, in the eigenstate basis charge fluctuations are off-diagonal
at this working point.
Therefore projecting the coupled boxes Hamiltonian \eref{eq:Noiseless-SingleQubit-H}
into the computational subspace $\{\ket{i}_1\otimes \ket{j}_2\}$ - $i,j \in \{+,-\}$
we get
\begin{eqnarray}
\fl  \tilde{\cal H}=\tilde{\cal H}_0 \,+\, \delta\tilde{\cal H}
 \label{Projected_H} \\
  \tilde{\cal H}_0 = -\frac{\Omega}{2}\sigma_3^{(1)}\otimes \mathbb{I}^{(2)} \,-\,
            \frac{\Omega}{2}\mathbb{I}^{(1)}\otimes \sigma_3^{(2)}
            +\frac{\tilde{E}_{\mathrm{CC}}}{2}\sigma_1^{(1)} \otimes \sigma_1^{(2)}
 \label{Projected_H1}  
\\
  \delta\tilde	{\cal H}=
            -\frac{\tilde{X}_1}{2}\sigma_1^{(1)}\otimes\mathbb{I}^{(2)} \,-\,
            \frac{\tilde{X}_2}{2}\mathbb{I}^{(1)}\otimes \sigma_1^{(2)} \,,
  \label{Projected_H2}
 \end{eqnarray}
where we assume the two qubits are tuned at the same Bohr splitting $\Omega$,
whose typical value is of the order $10^{11}$ rad/s.  Here 
$\tilde{E}_{\mathrm{CC}}=2E_{\mathrm{CC}}\,q_{1,+-}\,q_{2,+-}$ and 
$\tilde{X}_{\alpha}=2X_{\alpha}\,q_{\alpha,+-}$, 
being $q_{\alpha,+-}=\bra{+}_\alpha \hat{q}_\alpha\ket{-}_\alpha$.

We remark that  this symmetric configuration
is hardly reachable in practice by fabrication accuracy only.
In the charge-phase implementation~\cite{Vion02} this can be achieved thanks to
the characteristic two-port design. The quantization is in fact based on a
split CPB connected to a large measurement Josephson junction,  
the phase across it, $\delta$, representing an additional control knob. 
The device presents a ``doubly'' protected point at $q_\mathrm{x}=1/2$ and $\delta=0$, 
which is a saddle point of the single-qubit energy splitting versus external parameters.
In the two-qubit setup resonance is achieved by slightly displacing 
one of the qubits from the phase protected point, but maintaining both qubits at the
charge protected point. 
This setup is therefore expected to be sensitive also to phase fluctuations and this
topic will be addressed elsewhere~\cite{Mastellone08}. 

The setup described by \eref{Projected_H} may in principle implement a iSWAP gate. This
is easily illustrated in its eigenstate basis, reported in \tref{tab:eigensystem0}
in terms of the dimensionless coupling strength $g=\tilde{E}_{\mathrm{CC}}/\Omega$.
\begin{table}[!ht]
\caption{
 \label{tab:eigensystem0} Eigenvalues and eigenvectors of $\mathcal{H}_0$.
  Here $\sin \varphi = g/(2\sqrt{1+g^2/4})$ and $\cos \varphi = -1/\sqrt{1+g^2/4}$.}
  \begin{indented}
  \item[]
  \begin{tabular}{ccc}
  \br
  $i$ & $\lambda^{(0)}_i$ & $\vert i \rangle$ \\ 
  \mr
    0 & \quad $- \sqrt{1 + g^2/4}$ & \qquad 
     $[-(\sin \varphi/2)\vert ++ \rangle+(\cos\varphi/2)\vert-- \rangle ]$ \\
    1 & $-g/2$ & $(- \vert +- \rangle + \vert -+ \rangle )/\sqrt{2}$ \\
    2 & $g/2$ & $( \vert +- \rangle + \vert -+ \rangle )/\sqrt{2}$ \\
    3 & \quad $\sqrt{1 + g^2/4}$ & \qquad 
     $[(\cos \varphi/2)\vert ++ \rangle+(\sin \varphi/2)\vert -- \rangle ]$ \\
  \br
  \end{tabular}
  \end{indented}
\end{table}
In the absence of fluctuations, the two-qubit 
Hilbert space is factorized in two subspaces spanned by pairs of computational states. 
In particular, the system prepared in the state $\ket{+-}$, freely evolves inside the 
subspace spanned by $\{\ket{+-},\ket{-+}\}$, reaching the entangled state 
$(\ket{+-}+\rmi\ket{-+})/\sqrt{2}$ at time  
$\bar \tau= \bar t \Omega = \pi/2g$. States $\{ \ket{1},\ket{2} \}$ generate
the so called SWAP subspace,
whereas we refer to Z subspace as the one spanned by 
$\{\ket{0}, \ket{3} \}$~\cite{Mastellone08} .

A numerical analysis has  shown that for typical values of parameters in 
charge-phase qubits, the SWAP-eigenvalues  are  more stable than single qubit splitting
with respect to uncorrelated gate charge fluctuations~\cite{Mastellone_EPJ}.
As a consequence, for sufficiently small-amplitude low-frequency charge fluctuations
the decay time for entangled states in the SWAP subspace is expected
to be longer than the single qubit dephasing time~\cite{Mastellone_EPJ}.
In the following we analyze how low frequency  correlations between charge
noise felt by each qubit may influence the dynamics in the SWAP subspace.

\subsection{Relevant dynamical quantities}
\label{section:Methods}

In the presence of low frequency fluctuations
the calibration of the device is unstable. As a result, the quantum dynamics of
the interacting qubits will depend on the measurement protocol, as already
observed for single qubit gates. Ideal quantum protocols assume measurements
of individual members of an ensemble of identical (meaning that the preparation
is controlled) evolutions, defocusing occurring during time evolution.
In practice for solid state nanodevices several samples are collected during
an overall measurement time $t_\mathrm{m}$. Lack of control of the environment preparation
determines defocusing of the signal, analogous to inhomogeneous broadening
in NMR. The considerable initial reduction of the amplitude of the coherent oscillations
of single qubit gates affected by 1/f charge noise is due precisely to this effect
~\cite{Falci05,Ithier05}.
On the other hand the effect of low frequency noise on relaxation processes is 
negligible. Thus the system
dynamics can be treated in the adiabatic approximation for the low frequency
charge fluctuations.  Under this approximation scheme populations of the system 
eigenstates do not evolve. The relevant dynamical quantities are therefore the
off-diagonal elements of the system density matrix in the same basis.

The efficiency of the iSWAP protocol in the presence of 1/f spectra on both
qubits and 1/f cross-spectrum can be therefore extracted by evaluating a
single dynamical quantity, the coherence between the eigenstates of the
SWAP subspace. The two qubit density matrix in the presence of the
dimensionless classical stochastic processes $x_i(t) = X_i(t)/\Omega$
generally reads
\begin{equation}
 \rho(\tau) = \int \mathcal{D} [x_{1}(\tau')] \mathcal{D} [x_{2}(\tau')]
               P[x_{1}(\tau'),x_{2}(\tau')]
               \rho[\tau \vert x_{1}(\tau'),x_{2}(\tau')],
 \label{eq:noise-integral}
\end{equation}
where $\rho[\tau \vert x_{1}(\tau'),x_{2}(\tau')]$ is the system density 
matrix calculated for a given realization
$\{x_{1}(\tau'),x_{2}(\tau')\}$. The integration is over
over all possible realizations weighted by the probability density 
$P[x_{1}(\tau),x_{2}(\tau)]$.

We solved equation \eref{eq:noise-integral} numerically by generating the
independent random processes $n_1(\tau)$ and $n_2(\tau)$ and from them
the correlated processes $x_1(t)$ and $x_2(t)$, as illustrated 
in \sref{section:Noise-model}.
The Schr\"odinger equation related to the Hamiltonian \eref{Projected_H}
is numerically solved by a fourth order Runge-Kutta algorithm \cite{LandauPaez}, 
calculating the system dynamics
$\rho[\tau \vert x_{1}(\tau),x_{2}(\tau)]$. These operations are repeated 
to perform an average of $\rho[\tau \vert x_{1}(\tau),x_{2}(\tau)]$
over many ($\geq 10^4$) realizations of the stochastic processes.
Numerical simulations confirm that  in the presence of low
frequency noise (with $\gamma_\mathrm{M} < 10^{-2}\,\Omega$) 
transitions between the SWAP and Z subspaces can be neglected.
This further legitimates focusing on 
the coherence in the SWAP subspace which reads
\begin{eqnarray}
 \fl \bra{1}\rho(\tau)\ket{2} \equiv \rho_{12}(\tau) 
= \rho_{12}(0) \rme^{- \rmi g \tau}
  \rme^{-\rmi \Phi(\tau)} = \rho_{12}(0) \nonumber \\
   \!\!\!\!\!\!\!\!\!\cdot \int \mathcal{D} [x_1(\tau')] \mathcal{D} [x_2(\tau')]
  P[x_1(\tau'),x_2(\tau')]
  \exp \left [ \rmi \int_0^{\tau'} \rmd \tau'' \omega_{12}[x_1(\tau''),x_2(\tau'')] \right ],
\label{eq:slownoiseint}
\end{eqnarray}
where $\omega_{12}[x_1(\tau''),x_2(\tau'')]$ gives the noise renormalized 
splitting between states $\ket 1$ and $\ket 2$. 
The imaginary part of $\Phi(\tau)$ describes the decay of 
the entangled dynamics in the presence of adiabatic correlated noise.
Further insight can be obtained by approximating
\eref{eq:slownoiseint} to include the dominant inhomogeneous broadening
effect. This is performed by applying the static path approximation (SPA),
$x_\alpha(t)\equiv x_\alpha$, which accounts for the lack of control of the
device calibration via a statistical distributed gate charge at each
run of the measurement protocol.  
In the SPA the coherence \eref{eq:slownoiseint} reduces to the evaluation of
an ordinary two-variables integral 
\begin{equation}
 \rho_{12}(\tau) = \rho_{12}(0) 
 \int \rmd x_1 \,\rmd x_2 \,
  P(x_1,x_2) \,
  \exp \left [ \rmi \tau \omega_{12}(x_1,x_2)  \right]\,,
\label{eq:slownoiseintSPA}
\end{equation}
where $P(x_1,x_2)$ is the joint probability density function of the random 
variables $x_1$ and $x_2$~ \cite{Papoulis}. 
In the following we will use the notation $\langle f(x_1,x_2) \rangle$ to indicate
$\int \rmd x_1 \,\rmd x_2 \,P(x_1,x_2)\, f(x_1,x_2)$.

In the following Section we will analytically evaluate the coherence in the
SWAP subspace within the SPA in selected parameter regimes where  numerical simulations 
have confirmed its accuracy. A numerical analysis will be performed to 
estimate the decay of entanglement under more general conditions.

\section{Dephasing in the SWAP subspace: effects of correlations}
\label{section:Results}
The average in \eref{eq:slownoiseintSPA} is conveniently evaluated 
by performing the change of variables \eref{CorrelatedSP}. 
In fact, since the independent random processes $n_1(t)$ and $n_2(t)$ are generated 
from a large number of  TLFs, their initial values $n_\alpha$ are Gaussian distributed
\begin{equation}
        P(n_\alpha) = \frac{1}{\sqrt{2 \pi}\sigma}
        \exp [-n_\alpha^2/(2\sigma_{n_\alpha}^2)] \, .
\label{eq:gaussian}
\end{equation}
Clearly, 
$x_1$ and $x_2$ are two correlated Gaussian variables 
 whose joint probability density function is  (for $|\mu| <1$)~\cite{Papoulis}:
\begin{equation}
        P(x_1,x_2) = \frac{1}{2\pi\sigma^2\sqrt{1-\mu^2}}
        \exp \Big[-\frac{1}{2\sigma^2(1-\mu^2)}(x_1^2+x_2^2-2\mu x_1x_2)\Big] \, .
\label{eq:jointgaussian}
\end{equation}
The effective splitting in the SWAP subspace in the presence of charge
fluctuations entering the average  \eref{eq:slownoiseintSPA}  may be evaluated 
by exact diagonalization of the Hamiltonian \eref{Projected_H}. The solution 
of the resulting  fourth order polynomial is rather lengthy so we do not report 
it here. Relevant features can be extracted by expanding the splitting
up to fourth order in $x_\alpha$ and keeping the dominant terms in the
coupling strength $g$~\cite{Mastellone08}
\begin{equation}
        \omega_{12}
	(x_1,x_2) \approx  g - \frac{g}{2}(x_1^2 + x_2^2)
        + \frac{1}{8 g}(x_1^2-x_2^2)^2
	\equiv g \,+\, \delta\omega_{12}(x_1,x_2)\, .
\label{eq:swap-series-split}
\end{equation}
This expansion  suggests that 
the system behavior  depends on the relative weight of the amplitude of the 
noise, measured in the SPA by the standard deviation $\sigma$ entering \eref{eq:jointgaussian}, 
and the strength of the interaction between the qubits, $g$. In the following we consider
separately the two regimes of ``weak'' and ``strong'' amplitude noise,
$\sigma <g$ and $\sigma > g$ respectively. 
We remark that the nonmonotonous dependence of the splitting on $x_\alpha$ 
may lead to almost degeneracy between the renormalized levels of 
the SWAP subspace. This effect may be relevant when the interplay of
low and high frequency components is considered~\cite{Mastellone08}.

\begin{figure}[!t]
\centering
\includegraphics[width=80mm,height=53mm]{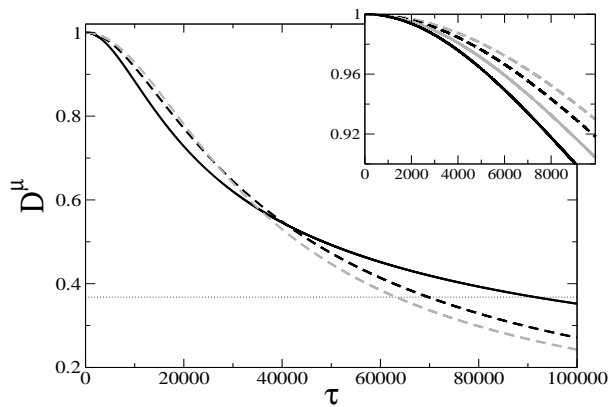}
\caption{Dephasing factor given by \eref{eq:DecayfactorWeakRegime} for different 
values of the correlation coefficient $\mu$ for ``weak'' amplitude noise,
$\sigma=0.02 < g=0.1$.
The dashed grey curve refers to the uncorrelated case $\mu=0$, 
the dashed black to $\mu=0.50$, the thick black 
 refers to $\mu=1.00$.
The crossing of each curve with the dotted horizontal line at   
$\rme^{-1}$ identifies to the estimated  
 dephasing time relative to each value of $\mu$, $\tau_2(\mu)$.
Intersections with the curve corresponding to uncorrelated noise ($\mu=0$)
identifies the time $\tau^*(\mu)< \tau_2$.
 Inset: enlargement for short times. The thick grey line corresponds to $\mu=0.75$.
The validity of the SPA approximation has been checked against numerical
simulations for stochastic processes exhibiting a 1/f power spectrum in a range 
$[\gamma_\mathrm{m},\gamma_\mathrm{M}]=[1,10^6]$ s$^{-1}$ (not shown).}
\label{fig2}
\end{figure}

{\em Weak amplitude noise $\sigma \ll g$:}
In this regime \eref{eq:swap-series-split} 
can be approximated by keeping terms up to second order in $\sigma$ 
so that $\delta\omega_{12}^{\mathrm{w}}(x_1,x_2) = -\frac{g}{2}(x_1^2 + x_2^2)$.
The average
\begin{equation}
 \rho_{12}(\tau) = \rho_{12}(0) \,\rme^{\rmi g\tau }\,
 \big\langle \exp (-\rmi\tau \delta\omega_{12}^{\mathrm{w}})\big\rangle
\label{eq:slownoiseintSPA_calculus}
\end{equation}
can be easily evaluated and leads, for the dephasing factor
\begin{eqnarray}
 \fl D^{\mu}_{12}(\tau)
\,=\, \Bigg\vert \frac{\rho_{12}(\tau)}{\rho_{12}(0)}\Bigg\vert 
=  \left[ 1+ (g\sigma^2 (1- \mu)\tau)^2 \right]^{-1/4}
\times \left[ 1+ (g\sigma^2 (1 + \mu)\tau)^2 \right]^{-1/4} \, .
  \label{eq:DecayfactorWeakRegime}
\end{eqnarray}
The dephasing factor factorizes into two contributions having the form of
the decay of the single qubit coherence at protected point in the SPA~\cite{Falci05} with
standard deviations $\sigma \sqrt{1 \pm \mu}$. An analogous result has been found in ~\cite{Hu07}. 
\Eref{eq:DecayfactorWeakRegime} is shown in  \fref{fig2} for different values
of the correlation coefficient $\mu$.
For comparison the curve corresponding to independent noise sources acting 
on the two qubits  is also reported,
$ D^{\mu=0}_{12}(\tau)=\left[1+(g\sigma^2\tau)^2\right]^{-1/2}$. 
\begin{figure}
\centering
\includegraphics[width=80mm,height=53mm]{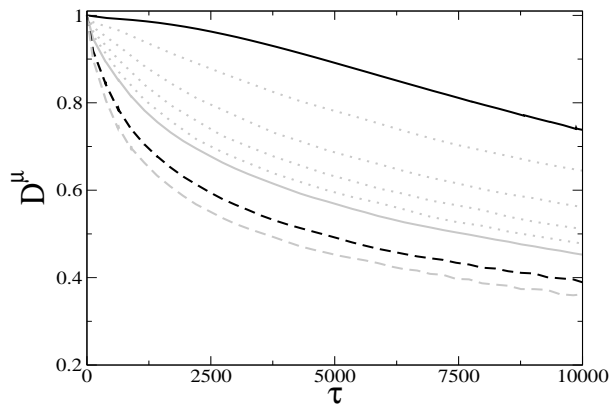}
\caption{
 Dephasing factor $D^{\mu}_{12}(\tau)$  for strong amplitude noise,
 $g=0.01 < \sigma = 0.08$. 
Curves correspond to different values of the correlation coefficient, 
from bottom to top  
$\mu \in \{0,0.50,0.75,0.8,0.85,0.90,0.95,1.00\}$.
Correlations improve the system performance despite the large noise amplitude.}
\label{fig4}
\end{figure}
Interestingly, 
at short times increasing the correlation coefficient induces a faster reduction
of the amplitude of coherent oscillations in the SWAP subspace.
This behavior crosses over to 
a regime where instead increasing the correlation coefficient slows down dephasing. This 
occurs  at times larger than $\tau^*$
identified by the condition 
$D^{\mu}_{12}(\tau^*) = D^{0}_{12}(\tau^*)$, which gives
$\tau^{*}(\mu) = 1/ (g\sigma^2\sqrt{1-\mu^2/2})$.
The crossover takes place at times shorter than the dephasing time where
$D^{\mu}_{12}(\tau_2) = e^{-1}$
\begin{equation}
 \tau_2(\mu)=\frac{\sqrt{-(1+\mu^2)+\sqrt{(1+\mu^2)^2+(e^4-1)(1-\mu^2)^2}}}
      {g\sigma^2(1-\mu^2)} \,> \, \tau^*(\mu) \, .
\label{eq:weaknoiseregime-decoherence-time}
\end{equation} 
We remark that, for quantum computing purposes it is crucial understanding 
the behavior at times shorter than the dephasing time.  For instance  
fault-tolerant quantum computation~\cite{kn:nielsen-chuang,kn:benenti-casati-strini},
i.e. implementing reliable quantum operations even in presence of errors, 
requires  errors to be maintained below a small threshold
(typically $\epsilon_{\mathrm{th}}\sim 10^{-4}\div 10^{-6}$). 
The  error of the iSWAP gate under investigation may be simply
estimated as (in the adiabatic approximation)  
\begin{equation}
    \epsilon\,=\, 1 - \bra{\psi}\rho(\tau)\ket{\psi}\,=\,
        1-\frac{1}{2}D^{\mu}_{12}(\tau) \,
\label{error}
\end{equation}
being $\ket{\psi}$ the iSWAP target state.
This leads, at the dephasing time $\tau_2$, to an error about $0.8$.
The correlation coefficient dependence of the dephasing time $\tau_2$~\cite{Hu07}    
does not reveal relevant features of the iSWAP gate operation occurring at initial time scales.
\begin{figure}
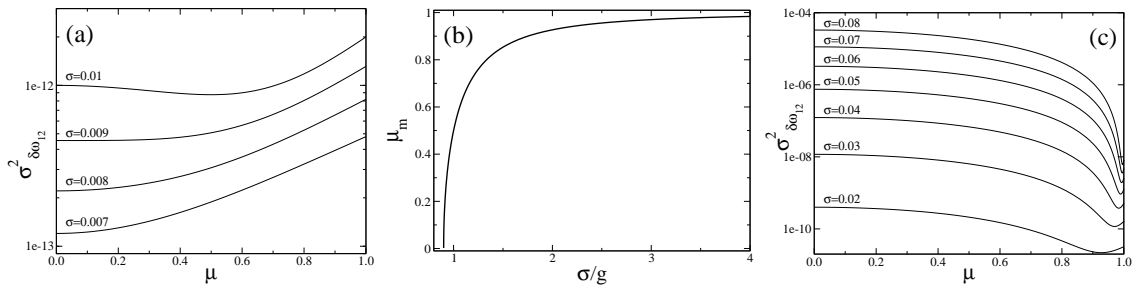

\centering
\includegraphics[width=49mm,height=37mm]{Figure/figure_5a}
\includegraphics[width=49mm,height=37mm]{Figure/figure_5b}
\includegraphics[width=49mm,height=37mm]{Figure/figure_5c}
 \caption{(a)
 Variance   \eref{eq:levelsplitting_variance-s} as a function $\mu$, for small
 amplitude noise $\sigma \lesssim g $.
(b) Correlation coefficient where
  the variance attains its minimum, 
  as function of the ratio $\sigma/g$. For $\sigma \geq 4g$ it is  $\mu_\mathrm{m} \simeq 1$. 
(c) \Eref{eq:levelsplitting_variance-s} as a function $\mu$ for $\sigma  \gg g$.
$\sigma^2_{\delta \omega_{12}}$ decreases by different orders of magnitudes
when  $\mu \geq 0.8$. In (a) and (c) $g=0.01$.}
\label{fig6}
\end{figure}

{\em Strong amplitude noise $\sigma \gg g$:}
For larger amplitudes of the noise, the dephasing factor displays a completely
different behavior. Numerical results are illustrated in \fref{fig4}. 
In this case increasing the correlation coefficient systematically {\em reduces} 
dephasing. In particular the
limiting case $\mu=1$, where the same noise simultaneously acts on both qubits,
turns out to be relatively weakly affected by the low frequency noise, despite of
its large amplitude.

The short time behavior and its dependence on the correlation coefficient
for weak and strong amplitude noise may be explained considering the renormalized
splitting dependence on the noise variables $x_\alpha$.
The level splitting itself in fact can be viewed as a random variable, with standard deviation
$\sigma_{\delta \omega_{12}}=\sqrt{\langle\delta \omega^2_{12}\rangle-
     \langle\delta \omega_{12}\rangle^2}$.
The short times gaussian approximation for the dephasing factor gives
\begin{equation}
 D^{\mu}_{12}(\tau)\simeq 1- \frac{1}{2}\sigma^2_{\delta \omega_{12}}\tau^2,
 \label{eq:ApproxDecayfactor}
\end{equation}
thus, at times short enough that $\sigma_{\delta \omega_{12}}\tau < 1$ 
(for data in \fref{fig2}, for $\tau<10^4$),
a larger deviation $\sigma_{\delta \omega_{12}}$ induces a larger 
dephasing, i.e. a smaller value for $D^{\mu}_{12}(\tau)$.
The variance  of
$\delta \omega_{12}$ reads (see appendix for details)
\begin{equation}
  \sigma^2_{\delta \omega_{12}}=
   \frac{\sigma^4}{g^2}
   \Big\{[(g^2-\sigma^2)^2+ \sigma^4]+
    \mu^2[(g^2+\sigma^2)^2-5\sigma^4]+
    2\mu^4\sigma^4]
   \Big\} \, .
 \label{eq:levelsplitting_variance-s}
\end{equation}
For $\sigma < g$, it reduces approximately to
 $ \sigma_{\delta \omega_{12}^{\mathrm{w}}}^2\,=\,g^2\sigma^4(1+\mu^2)$,
monotonically increasing with $\mu$,  as shown in \fref{fig6}a. 
This explains the stronger dephasing observed for small amplitude noise
at short times.
On the other hand, for $\sigma >g$, $\sigma^2_{\delta \omega_{12}}$
is nonmonotonic with a minimum at 
$
 \mu_{\mathrm{m}}=[1-\frac{1}{2}(\sigma/g)^{-2}-\frac{1}{4}(\sigma/g)^{-4}]^{1/2},
$
which rapidly approaches $1$ (\fref{fig6}b).
Thus for $\mu \to 1$ the splitting variance rapidly reaches its minimum
value, implying small dephasing at short times even for
large noise amplitudes, $\sigma \geq 4g$.

\subsection{Effects of higher frequencies and recalibration protocol}
\begin{figure}
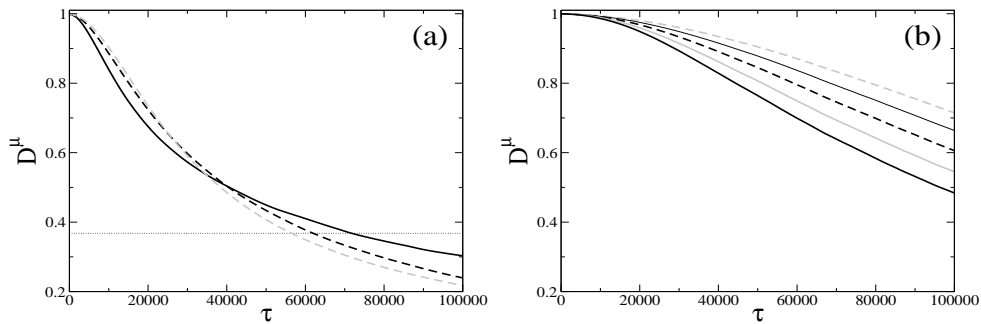

\centering
\includegraphics[width=64mm,height=42mm]{Figure/figure_6a}
\includegraphics[width=64mm,height=42mm]{Figure/figure_6b}
\caption{Effect of  1/f correlated noise  extending in  
$[\gamma_{\mathrm{m}},\gamma_{\mathrm{M}}]=[10^0, 10^9]$ s$^{-1}$ for weak noise amplitude $\sigma < g=0.1$
and different values of $\mu$. 
The dashed grey curve refers to $\mu=0$,  
 the dashed black to $\mu=0.50$, the thick black to $\mu=1.00$.
(a) Dephasing factor: correlations slightly increase dephasing at initial times and decrease it 
 at longer times. The gray horizontal line refers to $\rme^{-1}$. 
(b) Effect of a recalibration protocol:  increasing the correlation
coefficient increases dephasing. Thin black line corresponds to $\mu=0.25$, 
thick gray to $\mu=0.75$.}
\label{fig3}
\end{figure}

\begin{figure}
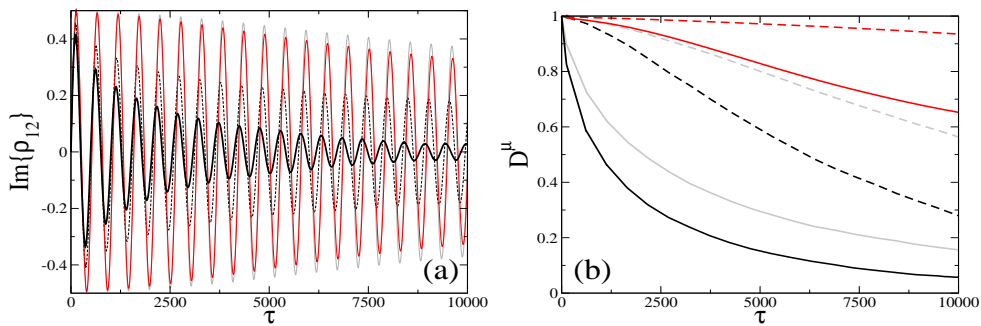

\centering
\includegraphics[width=64mm,height=42mm]{Figure/figure_7a}
\includegraphics[width=64mm,height=42mm]{Figure/figure_7b}
\caption{Effect of  1/f correlated noise  extending in  
$[\gamma_{\mathrm{m}},\gamma_{\mathrm{M}}]=[10^0, 10^9]$ s$^{-1}$ for strong noise amplitude $\sigma > g=0.01$
and different values of $\mu$. 
(a) Dephasing factor for $\mu=0$: black curves 
(light black for slow noise, tick black for 
low and high frequency noise); $\mu=1$: gray (slow noise) and red (slow plus
high frequency noise). (b) The effect of a recalibration protocol is shown by the
dashed curves, $\mu=0$ black, $\mu=0.75$ gray, $\mu=1$ red. 
In both protocols (a) and (b) increasing the correlation coefficient decreases dephasing.
}
\label{fig7}
\end{figure}

We now consider the effect of correlated 1/f noise extending to higher
frequencies maintaining the adiabaticity condition with respect to
the 
qubit splitting $\Omega$, i.e. we simulate fluctuations
leading to 1/f spectrum up to the cut-off frequency $\gamma_\mathrm{M} = 10^9$ s$^{-1}$.
The resulting dephasing in this case, in addition to the inhomogeneous
broadening mechanism, also originates from the dynamics of the
fluctuators during the time evolution. 
The dependence of the dephasing factor on the correlation coefficient $\mu$
has the same characteristics observed in the presence of low frequency components
only. 
The system still displays different behaviors depending on 
$\sigma$ being smaller or larger than $g$. This is illustrated in \fref{fig3}a
and \fref{fig7}a.
In particular the very low dephasing already observed 
when the qubits are affected by the same environment ($\mu=1$), 
persists also in the presence of high frequency
noise components (\fref{fig7}a).

In single qubit gates the inhomogeneous broadening effect may be sensibly 
reduced by a recalibration protocol resetting the initial value of the system 
polarization at each run of the measurement protocol~\cite{Falci05}.
Recalibration turns out to be effective on two-qubit gates also in the presence 
of correlations among the noise sources. 
Results shown in  \fref{fig3}b
and \fref{fig7}b have been obtained numerically by resetting the values of
$x_\alpha(0)$ at each realization of the stochastic processes $x_\alpha(t)$.
Interestingly, even if the effect of low frequency components is practically
eliminated by the recalibration procedure, the decay has a different dependence
on $\mu$ depending on $\sigma$.
In particular,  if $\sigma < g$ the larger is
the correlation coefficient, the faster the signal decays (\fref{fig3}b), 
if instead $\sigma > g$ stronger correlations correspond to slower decay
(\fref{fig7}b).

\section{Conclusions}
In the present paper we have introduced a phenomenological model for 1/f
correlated noise affecting a two-qubit gate in a fixed coupling scheme.
Our analysis is based on analytical results obtained within the adiabatic
approximation
and on the numerical solution of the stochastic Schr\"odinger equation.

Due to the nonmonotonicity of the renormalized splitting in the SWAP
subspace, the entangled dynamics sensitively depends on the ratio 
$\sigma/g$ between the amplitude of the low frequency noise and the qubits 
coupling strength.
For small amplitude noise, correlations increase dephasing at the relevant
short times scales (smaller than the dephasing time). On the other hand,
under strong amplitude noise, an increasing degree of correlations between 
noise sources acting on the two qubits always leads to reduced dephasing.
Our numerical analysis has shown that the above features hold true for
adiabatic 1/f noise extending up to frequencies $10^9$ s$^{-1}$ about
two order of magnitudes smaller that the qubit Bohr frequencies.

We remark that the above results on the reduced dynamics in 
the SWAP subspace apply also to diferent two-qubit gates involving the same 
states, at least as long as the adiabatic approximation 
holds true. The performance of two qubit gates involving states of the
Z subspace, like the c-NOT gate, might be reduced in view of the larger
sensitivity of the Z subspace to low-frequency charge noise~\cite{Mastellone_EPJ}. 

We have analyzed  the possibility to reduce the effects of low-frequency
correlated noise by a open-loop  recalibration protocol
of the two-qubit gate.
Despite counteracting the inhomogeneous broadening effect,
the efficiency of the protocol still depends on the value of  $\sigma/g$,
the maximum efficiency occurring for small amplitude uncorrelated noise
($\sigma < g$ and $\mu=0$),
or for strong amplitude correlated noise ($\sigma > g$ and $\mu=1$).

The observed reduced sensitivity of the SWAP subspace to correlated strong amplitude noise 
might suggest exploiting this subspace to reliably encode a single qubit, in the same 
spirit of Decoherence Free Subspaces (DFS)~\cite{Zanardi98}.
The possibility of avoiding errors due to correlated noise by encoding in DFS 
has indeed been recently discussed for superconducting qubits in~\cite{You05,Zhou04,Storcz05}.
This strategy rigorously applies to the pure dephasing regime where a DFS subspace
exits for collective noise.
In the situation analysed in the present article however the SWAP subspace is not 
rigorously a DFS. 
In fact, in the presence of collective noise ($\mu=1$)
the interaction Hamiltonian  \eref{Projected_H2} reads 
$-\frac{1}{2}\big(\sigma_1^{(1)}\otimes\mathbb{I}^{(2)} \,+\,
\mathbb{I}^{(1)}\otimes \sigma_1^{(2)} \big)\tilde X$. 
The system operator entering this coupling term has two degenerate eigenstates
which do not span the SWAP subspace. In addition, 
for finite values of $g$, the system Hamiltonian \eref{Projected_H1} does not leave invariant 
the subspace spanned by the degenerate eigenstates, as DFS should require~\cite{LidarWhaley03}.

\ack{
We thank Giuliano Benenti for useful discussions.
We acknowledge support from the EU-EuroSQIP (IST-3-015708-IP) and
MIUR-PRIN2005 (2005022977).
}

\appendix
\section{Moments entering  $\sigma_{\delta \omega_{12}}$}
From \eref{eq:jointgaussian} it comes out that the marginal probability density 
function of $x_\alpha$ is a Gaussian function 
with standard deviation $\sigma$ and zero average value. 
Then:
\begin{equation}
   \langle x_\alpha^{2n} \rangle = \frac{(2n)!}{2^n n!}\sigma^{2n}  \qquad \textrm{and} \qquad
   \langle x_\alpha^{2n+1} \rangle = 0,
\label{moments}
\end{equation}
see Ref.~\cite{Gardiner}.
Evaluation of the variance of the splitting fluctuations $\sigma_{\delta \omega_{12}}$
requires knowledge of the following  mixed moments:
\begin{eqnarray}
   \langle x_1^2x_2^2 \rangle\,&=&\, \sigma^4(1+2\mu^2) \nonumber \\
   \langle x_1^2x_2^4 \rangle\,&=&\, 3\sigma^6(1+4\mu^2)\nonumber \\   
   \langle x_1^2x_2^6 \rangle\,&=&\, 15 \sigma^8(1+6\mu^2)\nonumber \\   
   \langle x_1^4x_2^4 \rangle\,&=&\, \sigma^8(9+72\mu^2+24\mu^4) 
\label{mixedmomenta}
\end{eqnarray}
which  directly follow from \eref{CorrelatedSP} by using \eref{moments}.

\end{document}